\documentclass[fleqn,usenatbib]{mnras}

\usepackage{newtxtext,newtxmath}

\usepackage[T1]{fontenc}

\DeclareRobustCommand{\VAN}[3]{#2}
\let\VANthebibliography\thebibliography
\def\thebibliography{\DeclareRobustCommand{\VAN}[3]{##3}\VANthebibliography}

\usepackage{graphicx}	
\usepackage{amsmath}	
\usepackage{longtable}
\usepackage{graphicx}   
\usepackage{amsmath}    
\usepackage{dirtytalk}
\usepackage[normalem]{ulem}
\usepackage[table]{xcolor}
\usepackage{siunitx}
\usepackage{natbib}
\usepackage{hyperref}
\usepackage{breqn}
\usepackage{bm}

\DeclareUnicodeCharacter{2212}{-}

\title[X-ray spectral and timing analysis of NGC 1068]{X-ray spectral and timing analysis of the Compton Thick Seyfert 2 galaxy NGC 1068}

\author[I. Pal et al.]{
Indrani Pal,$^{1}$\thanks{E-mail: indrani.pal@iiap.res.in}
C. S. Stalin,$^{1}$
M. L. Parker,$^{2}$
Vivek. K. Agrawal $^{3}$ and
S. Marchesi $^{4,5}$
\\
$^{1}$Indian Institute of Astrophysics, Bangalore, India  \\
$^{2}$Institute of Astronomy, Madingley Road, Cambridge, CB3 OHA, UK \\
$^{3}$Space Astronomy Group, ISITE Campus, U R Rao Satellite Centre Bengaluru \\
$^{4}$INAF - Osservatorio di Astrofisica e Scienza dello Spazio di Bologna, Via Piero Gobetti, 93/3, 40129, Bologna, Italy \\
$^{5}$Department of Physics and Astronomy, Clemson University, Kinard Lab of Physics, Clemson, SC 29634, USA \\
}

\date{Accepted XXX. Received YYY; in original form ZZZ}

\pubyear{2022}

\begin{document}
\label{firstpage}
\pagerange{\pageref{firstpage}--\pageref{lastpage}}
\maketitle

\begin{abstract}
We present the timing and spectral analysis of the Compton Thick Seyfert 2 active galactic nuclei NGC 1068 observed using {\it NuSTAR} and {\it XMM-Newton}. In this work for the first time we calculated the coronal temperature ($\rm{kT_{e}}$) of the source and checked for its variation between the epochs if any. The data analysed in this work comprised of (a) eight epochs of observations with {\it NuSTAR} carried out during the period December 2012 to November 2017, and, (b) six epochs of observations with {\it XMM-Newton} carried out during July 2000 to February 2015. From timing analysis of the {\it NuSTAR} observations, we found the source not to show any variations in the soft band. However, on examination of the flux at energies beyond 20 keV, during August 2014 and August 2017 the source was brighter by about 20\% and 30\% respectively compared to the mean flux of the three 2012 {\it NuSTAR} observations as in agreement with earlier results in literature. From an analysis of {\it XMM-Newton} data we found no variation in the hard band (2 $-$ 4 keV) between epochs as well as within epochs. In the soft band (0.2 $-$ 2 keV), while the source was found to be not variable within epochs, it was found to be brighter in epoch B relative to epoch A. By fitting physical models we determined $\rm{kT_{e}}$ to range between  8.46$^{+0.39}_{-0.66}$ keV and 9.13$^{+0.63}_{-0.98}$ keV. From our analysis, we conclude that we found no variation of $\rm{kT_{e}}$ in the source.
\end{abstract}




\section{Introduction}

The standard orientation based unification model of active galactic nuclei (AGN;
\citealt{1993ARA&A..31..473A,1995PASP..107..803U}) classifies the Seyfert 
category of AGN  into two types, namely Seyfert 1 and Seyfert 2 galaxies,  based on the 
orientation of the viewing angle.  According to this model, the observational
difference between Seyfert 1 and Seyfert 2 galaxies is explained due to the
inclination of the line of sight with respect to the dusty torus in them. 
Seyfert 1 galaxies are those that are viewed at lower inclination angles and 
Seyfert 2 galaxies are viewed at higher inclination, with the central region in 
them  completely blocked by the dusty molecular torus that surrounds the broad 
line region (BLR). The detection of hidden BLR was first reported in NGC 1068, 
which forms the discovery of the first Type 2 AGN. This was  based on  
spectro-polarimetric observations, that revealed the presence of broad emission 
lines in polarized light \citep{1985ApJ...297..621A}. X-ray observations too 
provide evidence on  the presence of the obscuring torus in AGN 
\citep{1991PASJ...43..195A} with large X-ray column densities seen in  
Seyfert 2 galaxies.

\begin{figure}
	\centering
	\begin{minipage}{1.05\columnwidth}
		\centering
		\includegraphics[width=\textwidth]{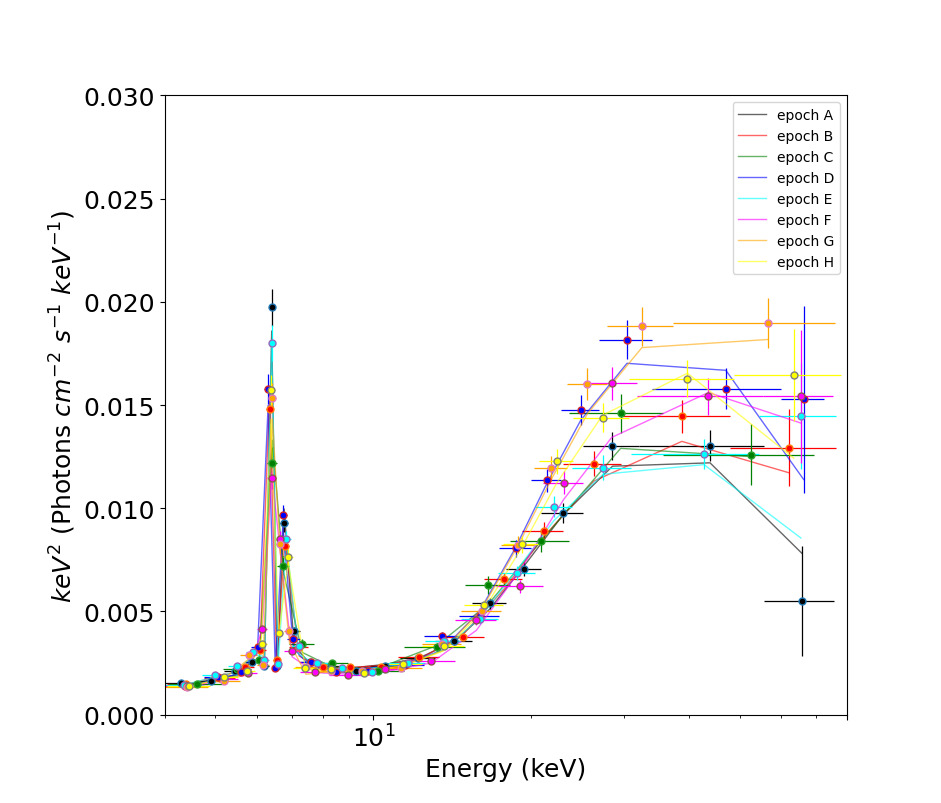}
		\end{minipage}%
\caption{The eight epochs of {\it NuSTAR} FPMA spectra plotted together} \label{figure-1}		
\end{figure}

AGN emit across wavelengths, and X-ray emission has been observed from  all 
categories of AGN \citep{1993ARA&A..31..717M}. However, a large fraction of AGN 
are obscured in X-rays \citep{2012AdAst2012E..17B, 2017NatAs...1..679R}. The 
obscured AGN are further classified as Compton Thin and Compton Thick based on 
the density of the equivalent hydrogen column density ($\rm{N_H}$) along the 
line of sight. In a Compton Thick AGN, the central engine is surrounded by 
the heavily obscuring dust and gas with $\rm{N_H}$ $\geq$ $10^{24}$ $cm^{-2}$ 
embedded in a dusty torus that is located at few parsec ($\sim$ 0.1 $-$ 10 pc) 
from the central source \citep{2015ARA&A..53..365N}. In a Compton Thick 
Seyfert 2 galaxy, reflection from the torus produces a 
reflection hump at around 15$-$30 keV and reveals the presence of neutral 
Fe K$\alpha$ emission line at 6.4 keV with an equivalent width of $\sim$1 keV 
(\citealt{1994MNRAS.267..743G,2016MNRAS.456L..94M}). However, the nature of this 
obscuring material is not static. In many AGN the observed X-ray variability is 
likely caused due to the variation in the circumnuclear material 
\citep{2002ApJ...571..234R, 2016ApJ...831..145Y}. 

X-ray emission from AGN is believed to originate from regions close to
the central super massive black hole. Models of the X-ray emitting region
include the hot corona situated close to the vicinity of the accretion 
disk (\citealt{1991ApJ...380L..51H,1993ApJ...413..507H})
as well as the AGN relativistic jet that emanates along the black hole
rotation axis \citep{2006ARA&A..44..463H,2019ARA&A..57..467B}. It is likely that 
the contribution of each of these physical processes to the observed X-ray 
emission may differ among AGN.  The observed X-ray spectrum from AGN consists of 
many components such as (i) a power law component believed to be due to the 
inverse Compton scattering of the optical, ultra-violet accretion disk photons by 
the hot electrons in the corona (\citealt{1991ApJ...380L..51H,1993ApJ...413..507H}), 
(ii) soft excess at energies lesser than 1 keV, which could be due to a warm 
($kT_e$ = 1 keV) and optically thick ($\tau$ = 10 - 20) corona 
\citep{2004MNRAS.349L...7G,2018A&A...611A..59P}  or due to relativistically 
blurred reflection \citep{2019ApJ...871...88G}, (iii) reflection bump beyond 
few keV due to scattering of X-rays by the accretion 
disk or distant material \citep{1994MNRAS.267..743G} and (iv) the fluorescent
Fe K$\alpha$ line with equivalent width of $\sim$1 keV\citep{1994MNRAS.267..743G,2016MNRAS.456L..94M}. 
In spite of the numerous X-ray studies on AGN, the exact causes for the origin of 
X-ray emission in them is still not understood. This also includes the size, shape 
and location of the X-ray corona in AGN. 
Parameters that can put constraints on the nature of the X-ray corona
in AGN are the power law index ($\Gamma$) and the high energy cut-off ($E_{cut}$)
in the X-ray continuum. This $E_{cut}$ in the X-ray continuum is
related to the temperature of the Comptonizing electrons ($kT_e$) in the
hot corona via E$_{cut}$ = 2$-$3 $kT_e$ \citep{2001ApJ...556..716P}, however, 
there are reports for this relation to show deviation among AGN \citep{2014ApJ...783..106L, 2019A&A...630A.131M, 2022A&A...662A..78P}.

One of the constraints to get good estimate of E$_{cut}$ on a large sample
of AGN is the lack of high signal to noise ratio data at
energies beyond 50 keV, where the cut-off in the spectrum is defined. 
Though more measurements are now available from instruments such as 
{\it NuSTAR} \citep{2015MNRAS.451.4375F,2018MNRAS.481.4419B,2018ApJ...866..124K,2018A&A...614A..37T,2018MNRAS.480.1819R,2018ApJ...856..120R,2019MNRAS.484.5113R,2020ApJ...901..111K,2020ApJ...905...41B,2021A&A...655A..60A,2021MNRAS.506.4960H,2022arXiv220200895K}, it is important to increase such measurements on a larger number of sources, firstly to find good estimates of E$_{cut}$ and secondly to find
better constraints on the correlation of E$_{cut}$ with other
physical properties of the sources. Also, variation in the temperature of the 
corona is now known in few radio-quiet category of AGN 
\citep{2018ApJ...863...71Z,2020MNRAS.492.3041B,2021ApJ...921...46B,2021MNRAS.502...80K,2022MNRAS.tmp..417W,2022A&A...662A..78P}, however, it is
not clear if it is shown by all AGN. This is due to the lack of such studies
on many  AGN, mostly attributed to the paucity of homogeneous multi-epoch 
data on a large number of sources. Therfore it is of great importance to find more 
sources that are known to show variations in $kT_e$.

\begin{table}
	\centering
	\caption{Log of {\it NuSTAR} observations.}
	\label{table-1}
	\begin{tabular}{cccc}
\hline
OBSID & Epoch & Date & Exposure Time \\
      &       &      &  (secs)         \\
\hline
60002030002 & A & 2012-12-18 & 57850  \\
60002030004 & B & 2012-12-20 & 48556  \\ 
60002030006 & C & 2012-12-21 & 19461  \\
60002033002 & D & 2014-08-18 & 52055  \\
60002033004 & E & 2015-02-05 & 53685  \\
60302003002 & F & 2017-07-31 & 49979  \\
60302003004 & G & 2017-08-27 & 52549  \\
60302003006 & H & 2017-11-06 & 49691  \\
		\hline
	\end{tabular}
\end{table}

\begin{table}
	\centering
	\caption{Log of {\it XMM-Newton} observations.}
	\label{table-2}
\begin{tabular}{cccc}
\hline
OBSID & Epoch & Date & Exposure Time \\
      &       &      &  (secs)         \\
\hline
 0111200101  & A & 2000-07-29 & 42258  \\
 0111200102 &  B & 2000-07-30 & 46429   \\
 0740060201  & C & 2014-07-10 & 63997   \\
 0740060301  & D & 2014-07-18 & 57600   \\
 0740060401  & E & 2014-08-19 & 54000   \\
 0740060501  & F & 2015-02-03 & 54600   \\

		\hline
	\end{tabular}
\end{table}

NGC 1068 is one of the most studied Seyfert 2 galaxies. Situated at a 
redshift of $z$ = 0.0038 \citep{1999ApJS..121..287H}, it is powered by a black hole of mass
$M_{BH} = 1.6 \times 10^7 M_{\odot}$ \citep{2006A&A...455..173P}.  In X-rays the source is 
studied in the past \citep{2004A&A...414..155M, Bauer_2015, 2016MNRAS.456L..94M, 
2020MNRAS.492.3872Z}. This source was first observed by {\it Ginga} in the X-ray band 
and the observation revealed the presence of a broad neutral Fe K$\alpha$ line 
\citep{1989PASJ...41..731K} with an equivalent width of $\sim$ 1.3 keV. Later 
on, the {\it ASCA} observations resolved the Fe lines into neutral and ionized 
components \citep{1994PASJ...46L..71U, 1997MNRAS.289..443I}. It is known as
an emitter of high energy $\gamma$-ray radiation in the 
MeV$-$GeV range \citep{2020ApJS..247...33A}  and also has been reported as a neutrino 
source \citep{2020PhRvL.124e1103A}. In the past, the hard X-ray source spectrum 
was fitted with a two reflector model \citep{1997A&A...325L..13M, 1999MNRAS.310...10G, Bauer_2015}. 
The central engine of the source is found to be completely obscured by the dusty 
torus with a column density of $\rm{N_H}$ $\geq$ $10^{25}$ $cm^{-2}$ \citep{2000MNRAS.318..173M}, 
therefore, the observer can only see the scattered emission along the line of 
sight. This scattered emission is commonly thought to originate from two types 
of reflectors, the \say{cold} reflector component that arises from Compton 
scattering off the primary X-ray emission from the neutral circumnuclear material, 
while the second \say{warm} ionized reflector component that arises due to Compton 
scattering off the heavily ionized material that acts as the \say{mirror} of the 
primary emission \citep{2004A&A...414..155M}.

Using the multi epoch X-ray observations \cite{Bauer_2015} fitted the spectra 
of NGC 1068 using the two reflector model along with different line emission, 
radiative recombination continuum and the off nuclear point source emission. 
Using {\it XMM-Newton} and {\it NuSTAR} joint fit of the NGC 1068 high 
energy ($>$ 4 keV) spectra \cite{2016MNRAS.456L..94M} detected  excess flux 
in August 2014 observation above 20 keV by 32$\pm$6 \% with respect to the 
previous 2012 December observation and later February 2015 observation. This 
transient excess above 20 keV in NGC 1068 spectra was ascribed to a drop in the 
absorbing column density from $\rm{N_H}$ $>$ 8.5 $\times$ $10^{24}$ $cm^{-2}$ 
to (5.9$\pm$0.4) $\times$ $10^{24}$ $cm^{-2}$ in 2012 spectra. The authors first 
caught the source during this unrevealing period in which the obscured material 
moved temporarily from the line of sight and the source was found to be its 
highest flux state.  Recently, \cite{2020MNRAS.492.3872Z} presented the 
spectral analysis of the {\it NuSTAR} data taken between July 2017 and Feb 2018 
to check for spectral variability. From the varying column density found in 
the timescale of 1 to 6 months, the authors inferred the presence of the clumpy torus 
structure surrounding the source. Using $\it Swift-XRT$ data the authors also 
detected an ultra-luminous X-ray source at a distance of $\sim$2 kpc from the 
nuclear region of NGC 1068.

\begin{figure*}
	\centering
	\begin{minipage}{2.10\columnwidth}
		\centering
		\includegraphics[width=\textwidth]{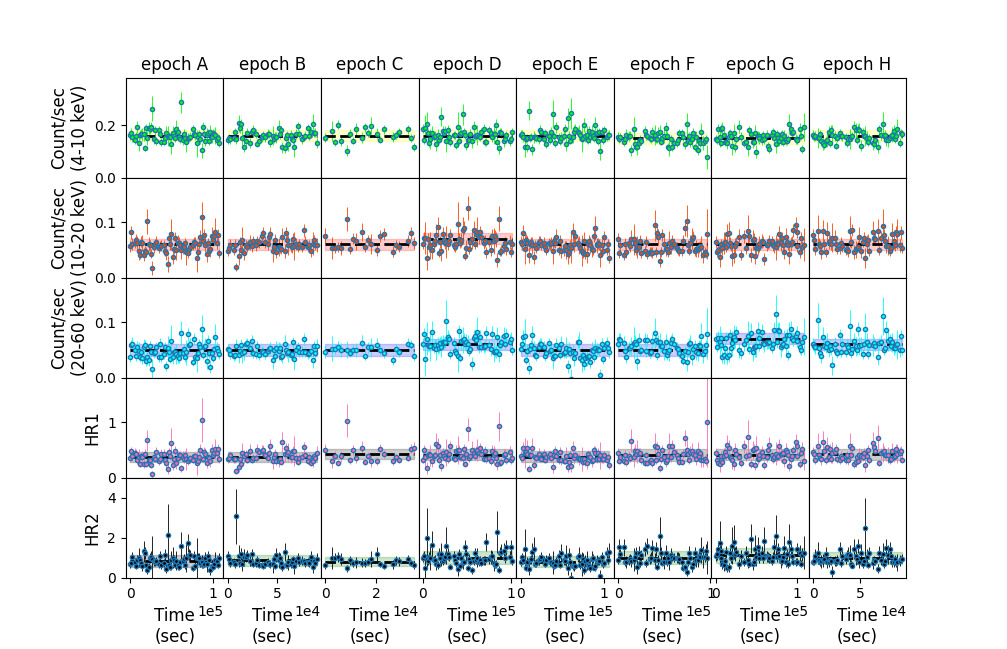}
		\end{minipage}%
\caption{The {\it NuSTAR} light curves of NGC 1068 in three energy bands, 
4$-$10 keV (first panel), 10$-$20 keV (second panel) and 20$-$60 keV (third panel). The HR1 and HR2 vs time are plotted in the last two panels. The black dashed lines are the mean of the count rate and HR. The shaded region in each panel is the mean errors in the count rate and HR.} \label{figure-2}		
\end{figure*}

\begin{figure*}
	\centering
	\begin{minipage}{1.05\columnwidth}
		\centering
		\includegraphics[width=\textwidth]{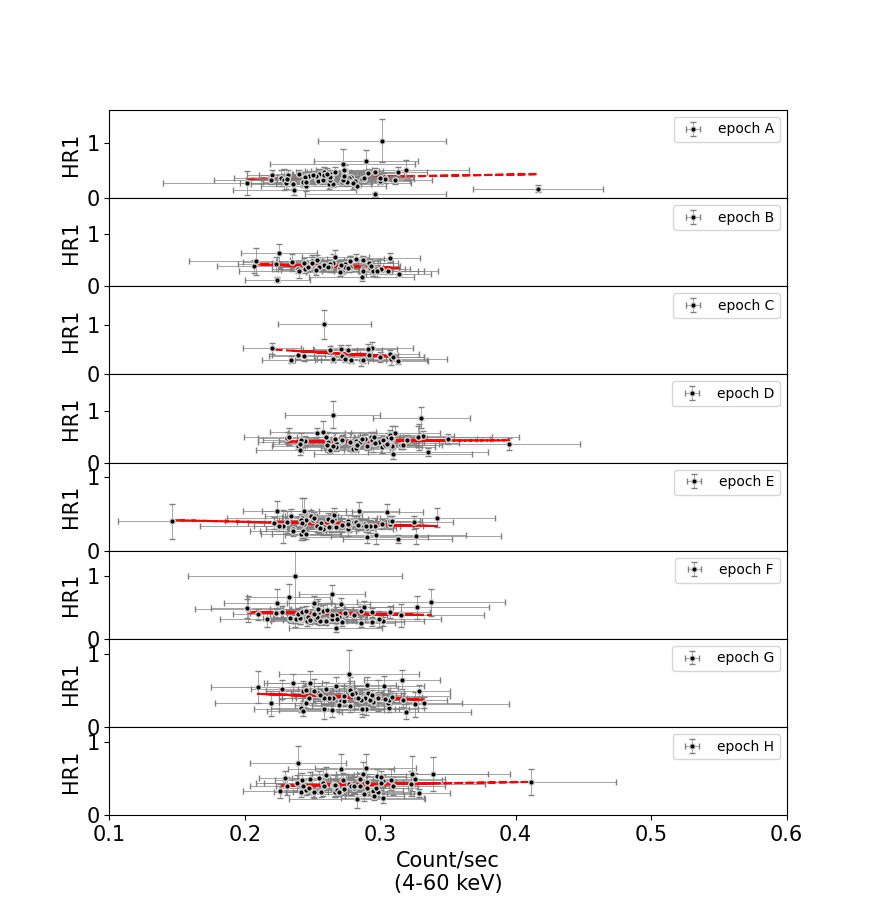}
		\end{minipage}%
	\begin{minipage}{1.05\columnwidth}
		\centering
		\includegraphics[width=\textwidth]{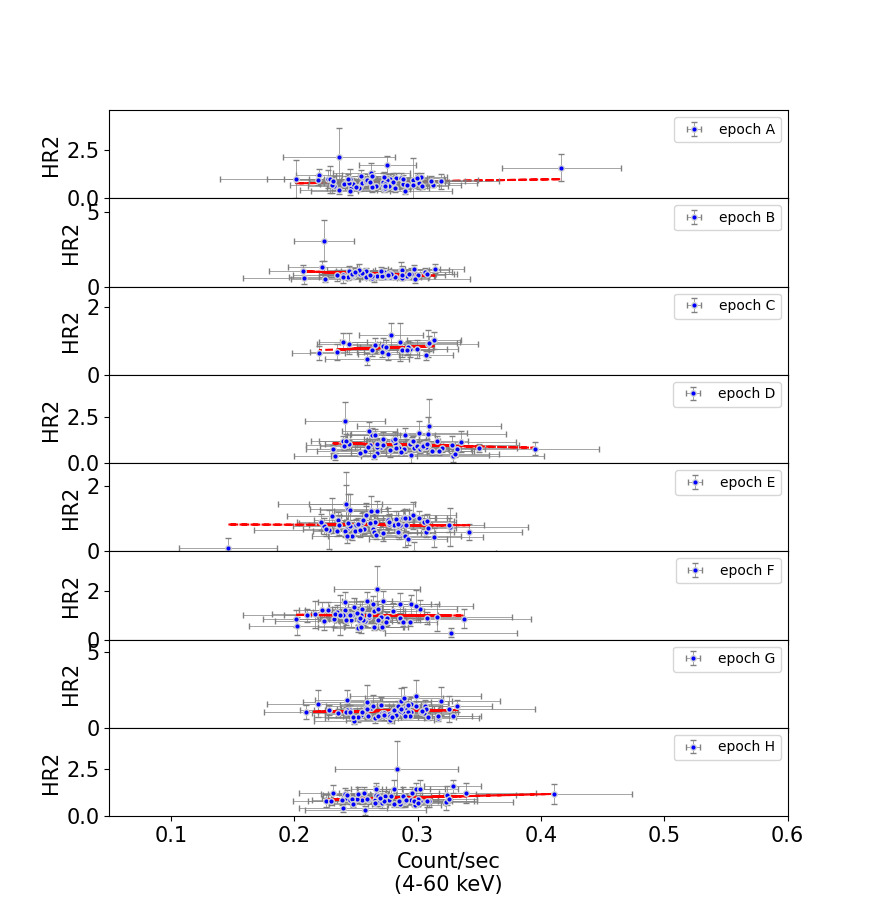}
		\end{minipage}
\caption{Left panel: The relation between HR1 and count rate in the 4$-$60 keV band. Right panel: 
The relation between HR2 and count rate in the 4$-$60 keV band. The red dashed lines in both panels are the linear least squares fit to the data.} \label{figure-3}		
\end{figure*}

\begin{figure*}
	\centering
	\begin{minipage}{2.10\columnwidth}
		\centering
		\includegraphics[width=\textwidth]{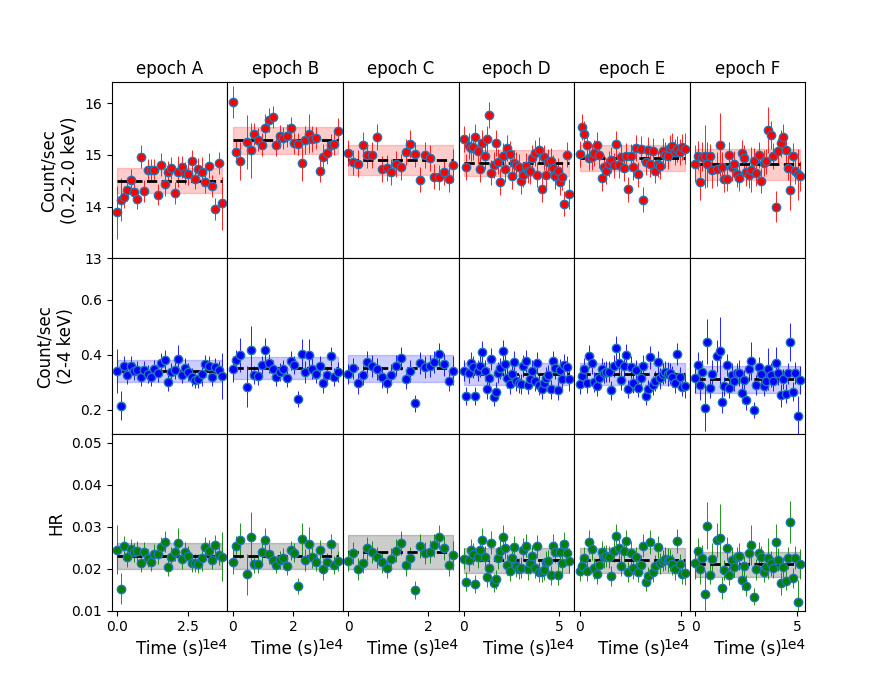}
		\end{minipage}%
\caption{{\it XMM-Newton EPIC PN} light curves of NGC 1068 in two energy bands, 
0.2$-$2 keV (top) and 2$-$4 keV (middle). The HR vs time is plotted in the bottom panel. The black dashed line and the shaded region in each panel is the mean value of counts/sec or HR and the corresponding errors respectively.} \label{figure-4}		
\end{figure*}

\begin{figure}
	\centering
	\begin{minipage}{1.05\columnwidth}
		\centering
		\includegraphics[width=\textwidth]{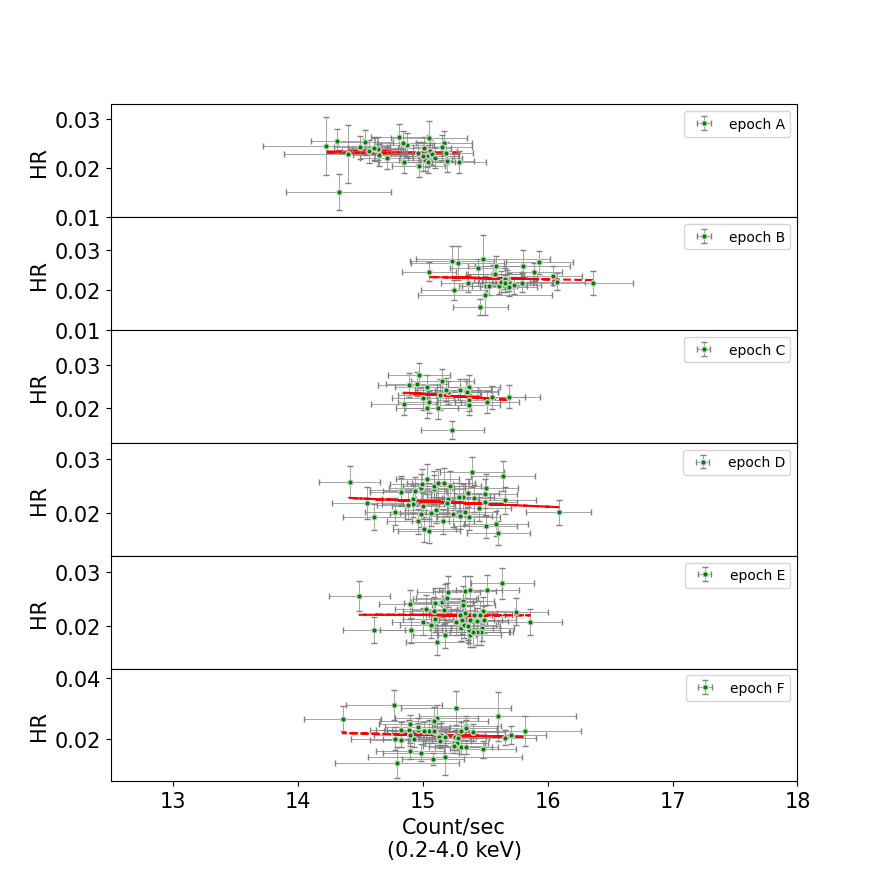}
		\end{minipage}%
\caption{The relation between HR and count rate in the 0.2$-$4 keV band. The red dashed lines in both panels are the linear least square fit to the data.} \label{figure-5}		
\end{figure}

Though the X-ray emission from NGC 1068 has been analysed in the past (\citealt{Bauer_2015}, \citealt{2016MNRAS.456L..94M}, \citealt{2020MNRAS.492.3872Z}), the source
has not been studied for variation in the temperature of the corona. \cite{Bauer_2015} from a joint fit of 2$-$195 keV data from different instruments 
reported a $\rm{E_{cut}}$ of 128$^{+115}_{-44}$ keV. Recently \cite{2021MNRAS.506.4960H} jointly fit the {\it XMM-Newton} (OBSID-0740060401) and {\it NuSTAR} (OBSID-60002033002) data and reported a $\rm{E_{cut}}$ of 28.4$^{+7.7}_{-4.0}$ keV. In this work, taking advantage of the multiple epochs of data available from {\it NuSTAR} (along with near simultaneous
XMM-Newton data at certain epochs of NuSTAR observations), we carried out for the first time an investigation of the variation in the temperature
of the corona if any. In this paper, we present results of our variability analysis of NGC 1068, from 
observations carried out by {\it NuSTAR} between 2012 and 2017. Also, we present 
our results on spectral analysis of the same {\it NuSTAR} data set in conjunction 
with observations from {\it XMM-Netwon}. The aim was to find the temperature of 
the corona in this source and its variation if any. The paper is organized as 
follows. In Section 2, we discuss the X-ray observations with {\it NuSTAR} and 
{\it XMM-Netwon} and the reduction of the data, the analysis is presented in Section 3, 
followed by the summary of the results in the final section.

\section{Observations and Data Reduction}
\subsection{{\it NuSTAR}}
Till now, NGC 1068 was observed by {\it NuSTAR} \citep{2013ApJ...770..103H} 
nine times with its two co-aligned telescopes with the focal plane modules A (FPMA) 
and B (FPMB) respectively.  In one of the epochs in the year 2018, an off-nuclear 
Ultra-Luminous X-ray source was detected \citep{2020MNRAS.492.3872Z}  at a distance 
of about $30''$ from the nuclear region of NGC 1068.  Barring this epoch, we considered eight epochs of data for this work. The details of these eight 
epochs of observations are given in Table \ref{table-1}. To visualize the spectral features of these observations the best fitted model-2 FPMA spectra (see section 4) are plotted together in Fig. \ref{figure-1}. 

We reduced the {\it NuSTAR} data in the 3$-$79 keV band using the standard {\it NuSTAR} 
data reduction software 
NuSTARDAS\footnote{https://heasarc.gsfc.nasa.gov/docs/nustar/analysis/nustar swguide.pdf} v1.9.7 
distributed by HEASARC within HEASoft v6.29. Considering the passage of the satellite 
through the South Atlantic Anomaly we selected SAACALC = \say{2}, SAAMODE \say{optimized} and also excluded the tentacle region. The calibrated, cleaned, and screened event 
files were generated by running {\tt nupipeline} task using the CALDB release 
20210701. To extract the source counts we chose a circular region of radius $50''$ 
centered on the source. Similarly, to extract the background counts,  we  
selected a circular region of the same radius away from the source on the same 
chip to avoid contamination from source photons. We then used the  {\tt nuproducts} 
task to generate energy spectra, response matrix files (RMFs) and auxiliary response 
files (ARFs), for both the hard X-ray detectors housed inside the corresponding 
focal plane modules FPMA and FPMB. 

\subsection{{\it XMM-Newton}}
NGC 1068 was observed by {\it XMM-Newton} for eight epochs during the year 2000 to 
2015. Within these we used six data sets taken between 2000 to 2015 for the timing analysis. For spectral analysis we used only three sets of data in 2014 due to their simultaneity with at least one epoch of {\it NuSTAR} observations.  

We chose to use only the {\it EPIC-PN} data for the extraction of source and background spectra. The log of the OBSIDs used in this work is given in Table \ref{table-2}. We used SAS v1.3 for the data reduction. Only single events (\say{PATTERN==0}) with quality flag=0 were selected. The event files were filtered to exclude background flares selected from time ranges where the 10–15 keV count rates in the PN camera exceeded 0.3 c/s. Source spectra were extracted from an annular region between the inner and outer radius of $15''$ and $30''$ centered on the nucleus. Background photons were selected from a source-free region of equal area on the same chip as the source. Here we note that for the source extraction, choosing a circular region of radius $30''$ produced pile up in the first two OBSIDs. However, pile up was not noticed in the other 4 epochs. To avoid pile up and to maintain uniformity in data reduction we chose to extract the source and background from an annular region for all the six epochs. We constructed RMFs and ARFs using the tasks {\it RMFGEN} and {\it ARFGEN} for each observation. 

\begin{table*}
\caption{Mean count-rate and the mean HR in different energy bands of NGC 1068 
obtained from the light curves (see Fig. \ref{figure-2}) and the results of the correlation between HR and the total count rate (see Fig. \ref{figure-3}).} \label{table-4}
\centering
\begin{tabular}{lcccccccccc}
\hline
OBSID & Epoch &  \multicolumn{3}{c}{Mean count rate} & \multicolumn{2}{c}{Mean HR} & \multicolumn{2}{c}{HR1} & \multicolumn{2}{c}{HR2} \\
      & &  4$-$ 10 keV & 10$-$20 keV & 20$-$60 keV & HR1  & HR2 & r & p & r & p \\
\hline
60002030002 & A & 0.16 $\pm$ 0.02 & 0.06 $\pm$ 0.01  & 0.05 $\pm$ 0.01  & 0.37 $\pm$ 0.09 & 0.85 $\pm$ 0.30 & 0.10 & 0.43 & 0.10 & 0.40 \\ 
60002030004 & B & 0.16 $\pm$ 0.02 & 0.06 $\pm$ 0.01  & 0.05 $\pm$ 0.01  & 0.38 $\pm$ 0.09 & 0.86 $\pm$ 0.27 & -0.18 & 0.20 & -0.18 & 0.19 \\ 
60002030006 & C & 0.16 $\pm$ 0.02 & 0.06 $\pm$ 0.01  & 0.05 $\pm$ 0.01  & 0.42 $\pm$ 0.09 & 0.80 $\pm$ 0.24 & -0.25 & 0.26 & 0.19 & 0.38 \\ 
60002033002 & D & 0.16 $\pm$ 0.02 & 0.07 $\pm$ 0.01  & 0.06 $\pm$ 0.01  & 0.41 $\pm$ 0.10 & 0.99 $\pm$ 0.32 & 0.04 & 0.78 & -0.12 & 0.34 \\ 
60002033004 & E & 0.16 $\pm$ 0.02 & 0.06 $\pm$ 0.01  & 0.05 $\pm$ 0.01  & 0.38 $\pm$ 0.10 & 0.80 $\pm$ 0.27 & -0.16 & 0.21 & -0.01 & 0.92  \\ 
60302003002 & F & 0.15 $\pm$ 0.02 & 0.06 $\pm$ 0.01  & 0.05 $\pm$ 0.01  & 0.41 $\pm$ 0.11 & 1.00 $\pm$ 0.32 & -0.07 & 0.58 & -0.02 & 0.90 \\ 
60302003004 & G & 0.15 $\pm$ 0.02 & 0.06 $\pm$ 0.01  & 0.07 $\pm$ 0.01  & 0.41 $\pm$ 0.10 & 1.12 $\pm$ 0.37 & -0.16 & 0.22 & 0.07 & 0.57 \\ 
60302003006 & H & 0.16 $\pm$ 0.02 & 0.06 $\pm$ 0.01  & 0.06 $\pm$ 0.01  & 0.42 $\pm$ 0.10 & 0.96 $\pm$ 0.30 & 0.09 & 0.52 & 0.16 & 0.22  \\ 
\hline
\end{tabular}
\end{table*}

\section{Timing Analysis}
\subsection{{\it NuSTAR}}
For timing analysis of the source, we utilized the data from {\it NuSTAR} 
and generated the background subtracted light curves with multiple corrections (e.g. bad pixel, livetime etc.) applied on the count rate in three energy bands,  namely, 
4$-$10 keV, 10$-$20 keV and 20$-$60 keV  respectively with 
a bin size of 1.2 ksec. The light curves in different energy
bands, along with variations in hardness ratios (HRs) are given
in Fig. \ref{figure-2}. To check for variability in the 
generated light curves we calculated the fractional root mean square
variability amplitude ($F_{var}$;\citealt{2002ApJ...568..610E,2003MNRAS.345.1271V}) 
for each epoch. $F_{var}$ is defined as
$F_{var} = \sqrt{\frac{V^2 - \overline{\sigma^2}}{\overline{x}^2}}$, 
where, $V^2 = \frac{1}{N-1} (x_i - \overline{x})^2$ is the sample variance 
and $\overline{\sigma^2} = \frac{1}{N} \sum \sigma_{i}^2$ is the mean square 
error in the flux measurements. Here, $x_i$ is the observed value in
counts per second, $\overline{x}$ is the arithmetic mean of the $x_i$ measurements,
and $\sigma_i$ is the error in each individual measurement. The error in 
$F_{var}$ was estimated following \cite{2003MNRAS.345.1271V}. For a binning choice of 1.2 ksec the calculated $F_{var}$ values indicate that the source is found not to show any significant variations within epochs. This is also evident in Fig. \ref{figure-2}. Shown by black dashed lines in Fig. \ref{figure-2} is the mean brightness of the source at each epoch determined from the light curves. These mean values are given in Table \ref{table-4}. From light curve analysis it is evident that the source has not shown variation in the soft band (4$-$10 keV and 10$-$20 keV) during the five years of data analysed in this work. However, variation is detected in the hard band (20$-$60 keV) (Fig. \ref{figure-2}). This is also very clear in Fig. \ref{figure-1}.

\begin{table*}
\caption{Results of the variability analysis in two energy bands of {\it XMM-Newton}} \label{table-6}
\centering
\begin{tabular}{lcccccc}
\hline
OBSID & Epoch & \multicolumn{2}{c}{Mean count rate} & Mean HR & r & p \\  
& & 0.2$-$2 keV & 2$-$4 keV &   &  &  \\
\hline
0111200101 & A & 14.50$\pm$0.24 & 0.34$\pm$0.04 & 0.023$\pm$0.003 & -0.03 & 0.88 \\
0111200201 & B & 15.28$\pm$0.26 & 0.35$\pm$0.04 & 0.023$\pm$0.003 & -0.06 & 0.76 \\
0740060201 & C & 14.86$\pm$0.25 & 0.34$\pm$0.04 & 0.023$\pm$0.003 & -0.16 & 0.45 \\
0740060301 & D & 14.85$\pm$0.25 & 0.33$\pm$0.04 & 0.022$\pm$0.003 & -0.11 & 0.45 \\
0740060401 & E & 14.94$\pm$0.25 & 0.33$\pm$0.04 & 0.022$\pm$0.003 & -0.01 & 0.96 \\
0740060501 & F & 14.82$\pm$0.30 & 0.31$\pm$0.05 & 0.021$\pm$0.003 & -0.06 & 0.71 \\
\hline
\end{tabular}
\end{table*}

In the two bottom panels of Figure \ref{figure-2}, we show the evolution of two hardness
ratios, namely HR1 and HR2 during the duration of the observations analysed
in this work. HR1 and HR2 are defined as: HR1 = C(10$-$20)/C(4$-$10) and HR2 = C(20$-$60)/C(10$-$20), where C(4$-$10), C(10$-$20) and
C(20$-$60) are the count rates in 4$-$10 keV, 10$-$20 keV, and 20$-$60 keV respectively. 
For each epoch, the mean hardness ratio is shown as a black dashed line in 
Figure \ref{figure-2} and the mean values are given in Table \ref{table-4}. As 
the errors are large, no variation in the hardness ratio of the source could be 
ascertained between epochs.  We also looked for correlation if any between the 
hardness ratios, HR1 and
HR2 against the broad band count rate in the 4$-$60 keV band with a time 
binning of 1.2 ksec. This is shown in Fig. \ref{figure-3}. Also, shown in the 
same figure are the linear least squares fit to the data. Calculated values of the Pearson's rank coefficient (r) and probability for no correlation (p) from the linear least square fit are given in Table \ref{table-4}. Analysing those values we found no variation of hardness ratios with the brightness of the source. 

\subsection{{\it XMM-Newton}}
Using the six epochs of {\it XMM-Newton EPIC PN} data from Table \ref{table-2} we generated the light curves in two energy bands, 0.2$-$2.0 keV and 2.0$-$4.0 keV using a binning size of 1.2 ksec. The light curves along with the variation of HR are shown in Fig. \ref{figure-4}. Here, HR is defined as the ratio of C(2.0$-$4.0) to C(0.2$-$2.0), where C(2.0$-$4.0) and C(0.2$-$2.0) are the count rates in 2.0$-$4.0 keV and 0.2$-$2.0 keV energy bands respectively. From $F_{var}$ analysis we found no significant variation within the epochs of observation. The black dashed lines in the first two panels of Fig. \ref{figure-4} are the mean values of the count rate in different energy bands. The mean values of the count rate  (see Table \ref{table-6}) indicate that in the soft band (0.2$-$2 keV) the source was in its brightest state in epoch B. There is thus variation in the soft band with the source being brighter in epoch B relative to epoch A. However, in the hard band we did not find any significant change in the source brightness between epochs. In the same Table \ref{table-6}, the constant values of mean HR between epochs also argues for no variation in brightness state of the source. In Fig. \ref{figure-5} HR is plotted against the total count rate in 0.2$-$4.0 keV band. The results of the linear least squares fit are given in Table \ref{table-6}. From the p values we conclude that no significant correlation between HR and the total count rate is found in NGC 1068.

\section{Spectral analysis}
In addition to characterizing the flux variability  of 
NGC 1068,  we also aimed in this work to investigate the variation in the 
temperature of the corona of the source. 
\subsection{NuSTAR only spectral fit}
To check for variation in the temperature of the corona in NGC 1068,  we first concentrated 
on the NuSTAR data alone. For that we fitted simultaneous FPMA/FPMB data for the eight epochs of observations available in the {\it NuSTAR} archive. To avoid the host galaxy contamination we used the {\it NuSTAR} data in the 4$-$79 keV energy band \citep{2016MNRAS.456L..94M}. For the spectral analysis, using XSPEC 
version 12.12.0 \citep{1996ASPC..101...17A}, we fitted the background subtracted spectra from FPMA and FPMB simultaneously (without combining them)  allowing the cross normalization factor to vary freely during spectral fits. The spectra were
binned to have minimum 25 counts/energy bin using the task {\it grppha}. To get 
an estimate of the  model parameters that best describe the observed data, we 
used the chi-square ($\chi^2$) statistics and for calculating the errors in the 
model parameters we used the $\chi^2$ = 2.71 criterion i.e. 90 \% confidence 
range in XSPEC.

\begin{figure*}
	\centering
	\begin{minipage}{1.05\columnwidth}
		\centering
		\includegraphics[width=\textwidth]{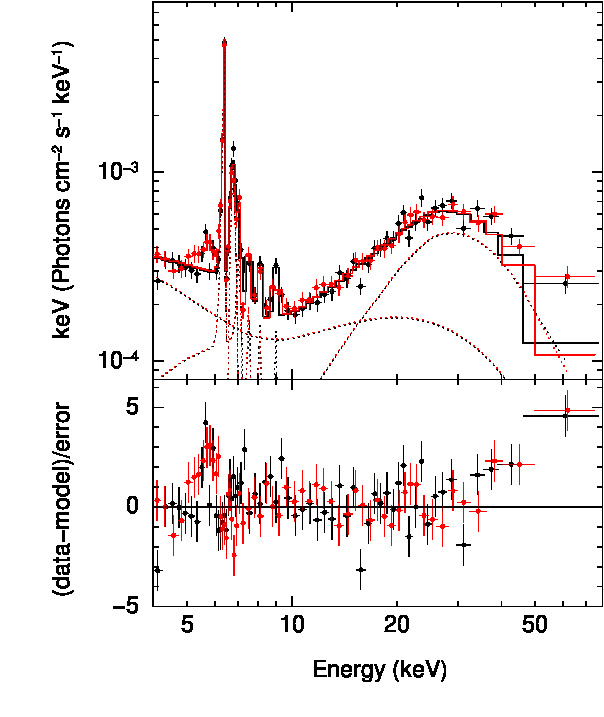}
		\end{minipage}%
	\begin{minipage}{1.05\columnwidth}
		\centering
		\includegraphics[width=\textwidth]{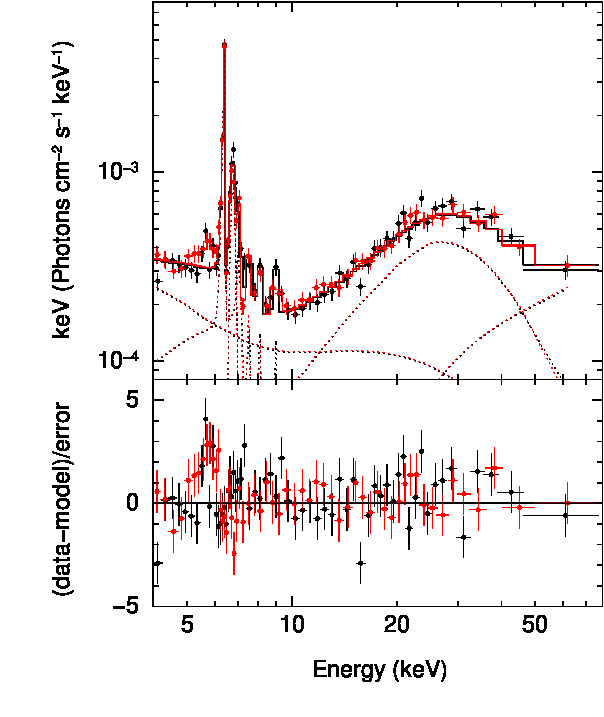}
		\end{minipage}
\caption{The best fit epoch G (with highest flux) unfolded spectra along with the data to model ratio using Model 1b (left panel) and Model 2b (right panel)}.\label{figure-7}		
\end{figure*}

\begin{figure*}
	\centering
	\begin{minipage}{1.05\columnwidth}
		\centering
		\includegraphics[width=\textwidth]{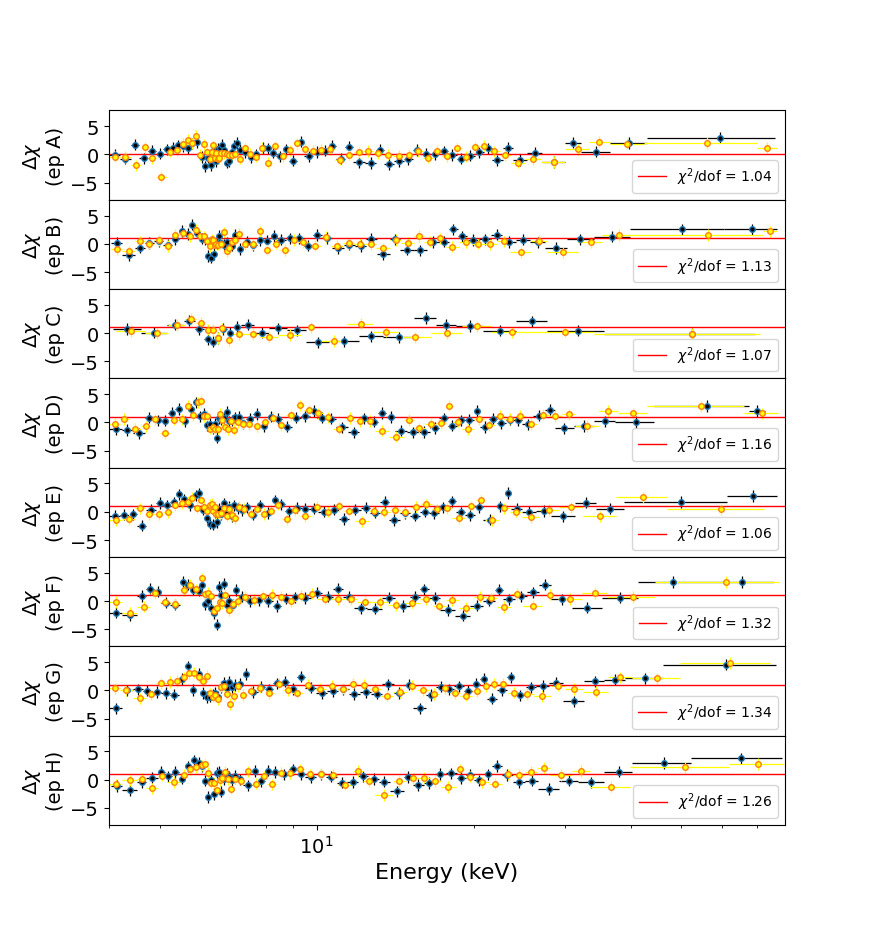}
		\end{minipage}%
	\begin{minipage}{1.05\columnwidth}
		\centering
		\includegraphics[width=\textwidth]{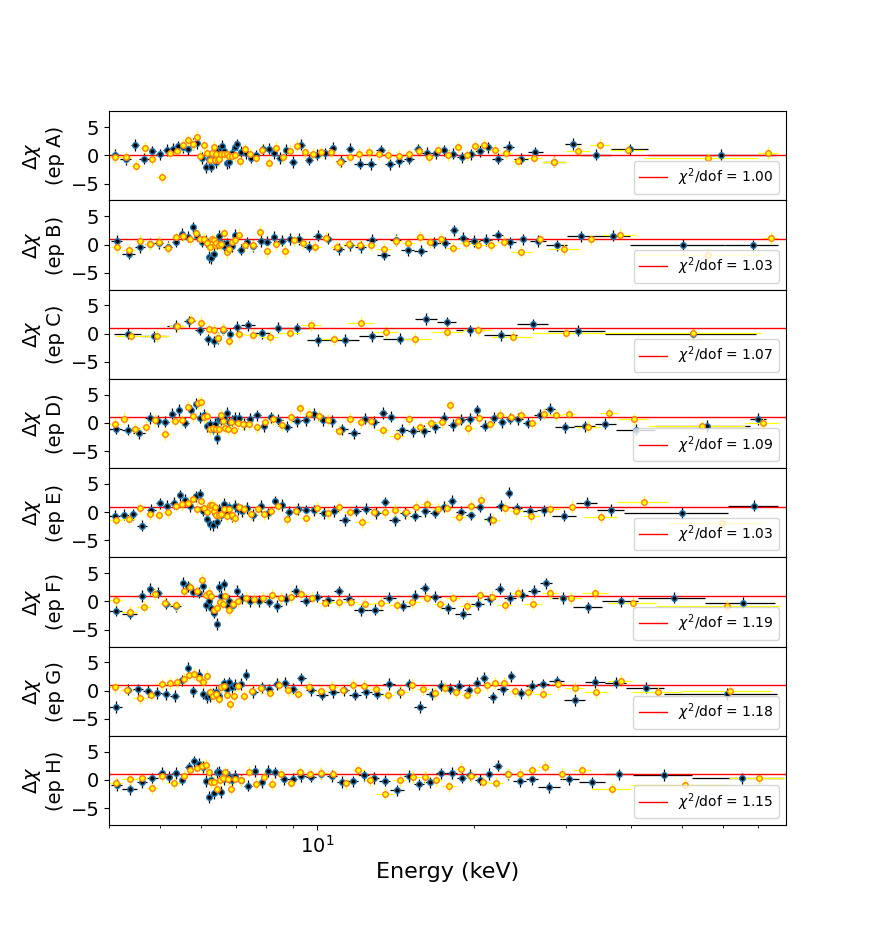}
		\end{minipage}
\caption{The data to model ratio of all eight epochs of {\it NuSTAR} observations using Model 1b (left panel) and Model 2b (right panel).} \label{figure-8}		
\end{figure*}

In all our model fits to the observed spectra, the {\it const} represents the cross calibration constant between the two focal plane modules FPMA and FPMB. To model the line of sight galactic absorption the {\it phabs} component was used and for this the neutral hydrogen column density ($\rm{N_H}$) was frozen to 3.32$\times$ $10^{20}$ atoms $cm^{-2}$ as obtained from \cite{2013MNRAS.431..394W}. To take into account the strong emission lines seen in the observed {\it NuSTAR} spectra, we used four {\it zgauss} components in XSPEC. In all the {\it zgauss} models  considered to fit the emission lines, the line energies and the normalization were kept free during the fitting while $\sigma$ was kept frozen to 0.1 keV. The redshift (z) for all the model components was kept fixed to 0.0038 \citep{2010A&A...518A..10V}. The inclination angle which is the angle the line of sight of the observer makes with the axis of the AGN ($i$) was fixed at 63$^{\circ}$ \citep{2004A&A...414..155M} in all the models.

\subsubsection{Model 1}
NGC 1068 has been extensively studied in the hard 
X-ray ($>$3 keV) band earlier \citep{1997A&A...325L..13M, 2004A&A...414..155M, 
Bauer_2015,2016MNRAS.456L..94M, 2020MNRAS.492.3872Z} mostly using the two 
component reflector model. The main essence of this model is to fit (a) the cold and distant reflector using {\it pexrav}/{\it pexmon}/({\it MYTS+MYTL}), and (b) the warm, ionized Compton-scattered component using {\it power-law/cutoff power-law} with Compton down scattering under the assumption that the electron temperature is much smaller than the photon energy ($m_{e}c^{2}$). Few Gaussian components were also used to model various neutral and ionized emission lines present in the source spectra. In this work too, to find the coronal temperature of the source, we modeled the source spectra using two different reflector components. 

\begin{enumerate}
\item The self consistent Comptonization model {\it xillverCP} \citep{2014ApJ...782...76G} that takes into account the cold and distant reflector, the neutral Fe k$\alpha$ ($\sim$ 6.4 keV) and Fe k$\beta$ ($\sim$ 7.06 keV) lines.

\item The warm ionized Compton scattered reflection using two models separately. At first, following \cite{Poutanen_1996}, a Compton scattered component ($f_{scat}$) for an arbitrary intrinsic continuum ($f_{intr}$). As an intrinsic continuum, we used a {\it power-law}, that was modified for Compton down scattering using equation (1) given in \cite{Poutanen_1996}. Secondly, the self consistent {\it xillverCP} model with a high ionization parameter ($log\xi$ = 4.7) to model the warm, ionized reflector. Here we note that fixing the ionization parameter to some other values ($log\xi$ = 3.0, 3.5 and 4.0) did not produce any significant change in the derived best fit values. It only self-consistently added few ionization lines in the model, but, the spectral shape remained unchanged. Using the Compton scattered component in place of a warm mirror may affect the spectra by adding curvature below 2 keV \citep{Bauer_2015}, but the spectral modeling of the {\it NuSTAR} data above 4 keV with or without inclusion of the Compton down scattering in the warm reflector did not produce any significant effect on the derived best fit values. We also arrived at the similar conclusion from using the {\it XMM-Newton} and {\it NuSTAR} data jointly below 4 keV. This is discussed in detail in Section 4.2.

\item Gaussian components to take care of the Fe
ionized ($\sim$ 6.57, 6.7 and 6.96 keV), Ni k$\alpha$ ($\sim$ 7.47 keV), Ni k$\beta$ ($\sim$ 8.23 keV) and Ni ionized ($\sim$ 7.83 keV) emission lines.
\end{enumerate}

In XSPEC, the two models used in the fitting of the spectra have the 
following form,

\begin{dmath}
Model 1a =   const*phabs*(f1*zpo+xillverCP+zgauss+zgauss+zgauss+zgauss)
\end{dmath}

and
\begin{dmath}
Model 1b =  const*phabs*(xillverCP_{warm}+xillverCP_{cold}+zgauss+zgauss+zgauss+zgauss)
\end{dmath}

Here, we note that, From the data to ratio plot we did not find any prominent residuals near the line emission regions, but, in all epochs we noticed residues at around 6.0 keV (see Fig. \ref{figure-7} and Fig. \ref{figure-8}). This feature at 6.0 keV has no physical origin but might appear in the {\it NuSTAR} data due to calibration issues \citep{2020MNRAS.492.3872Z}.

\noindent {\bf Model 1a:}
For the spectral fit with this model, we used the formula (f1) obtained from \cite{Poutanen_1996} to consider the Compton down scattering of the intrinsic continuum ({\it zpo}). Following \cite{Poutanen_1996},

\begin{equation}
    f1 \propto \tau_{sc}[1+\mu^2+xx_1(1-\mu)^2]
\end{equation}

In Equation 3, $x = h\nu/m_ec^2$ is the dimensionless photon energy
$\mu$ = cos i, x1 = x/[1- x(1-$\mu$)] and $\tau_{sc}$ is the Thompson optical
depth of the scattering material. We considered the constant of proportionality $\times$ $\tau_{sc}$ as an another constant (p1) and kept it as a free parameter in the spectral analysis. During spectral fits the photon index ($\Gamma$) and the normalization for the two reflectors were tied together. For the cold reflector we modeled only the reflection component by fixing the reflection fraction ($R$) to $-$1 throughout. The parameters that were kept free are the relative iron abundance ($AF_{e}$),  $\rm{kT_{e}}$ and p1. The constant, p1 was allowed to vary between 0.0 and 10.0 during the fit. To model the cold and neutral reflector the ionization parameter ($\xi$) was frozen to 1.0 (i.e log$\xi$ = 0). The best fit values obtained using Model 1a are given in Table \ref{table-8}.

\noindent {\bf Model 1b:} 
Here, we used {\it xillverCP} twice to model the warm and cold reflection respectively. For the warm and ionized reflector we used ${\it xillverCP_{warm}}$ by fixing the ionization parameter to its highest value ($log\xi = 4.7$) and for the cold and distant reflection (${\it xillverCP_{cold}}$) the reflector was considered as a neutral one with a  fixed $log\xi$ of 0.0. In the modelling of the source spectra using Model 1b we tied $\Gamma$ and $\rm{kT_{e}}$ of the two reflectors together. At first, the normalization for the two reflectors were tied together and the best fit produced a $\chi^2$ of 637 for 484 degrees of freedom for epoch A. We then modelled the epoch A spectrum by varying the two normalization independently and got an improved $\chi^2$ of 504 for 483 degrees of freedom. For the other epochs we therefore carried out the model fit by leaving the two normalization untied. For both the reflectors we fixed $R$ to $-$1 to consider the reflection components only. During the fitting $AF_{e}$ between the two reflectors were tied together. The best fit unfolded spectra along with the residues of the data to Model 1b fit to the epoch G spectra are given in the left panel of Fig. \ref{figure-7}. The best fit results of Model 1b are given in Table \ref{table-8}. For all the epochs the residuals of the fit are given in left panel of Fig. \ref{figure-8}.

\subsubsection{Model 2}
Following \cite{Bauer_2015} we then used the \say{leaky torus} model in which it is assumed that there is a finite probability for the primary emission to escape the medium without scattering or getting absorbed and partially punching through above 20 $-$ 30 keV. In a Compton-Thick AGN, the direct transmitted continuum if at all present, is not observable below $\sim$ 10 keV. In Model 2, with the two reflectors, this transmitted or the direct component was taken care of. We assumed that a direct transmitted intrinsic continuum ({\it zpo}) was attenuated by the line of sight Compton thick absorber with a column density of $\rm{N_{H}}$ = $10^{25}$ atoms $cm^{-2}$ and an inclination angle ($\theta_{incl}$) of $90^{\circ}$ (for an edge on torus). Here also, we used {\it xillverCP} with $log\xi$ = 0.0 to model the cold reflection, and, either {\it f1*zpo} \citep{Poutanen_1996} (Model 2a), or {\it xillverCP} with $log \xi$ = 4.7 (Model 2b) to take care of the warm and ionized reflection. In XSPEC the models take the following forms,

\begin{dmath}
Model 2a =  const*phabs*(zpo*MYTZ+f1*zpo+xillverCP++zgauss+zgauss+zgauss+zgauss)
\end{dmath}

and,

\begin{dmath}
Model 2b =  const*phabs*(zpo*MYTZ+xillverCP_{warm}+xillverCP_{cold}++zgauss+zgauss+zgauss+zgauss)
\end{dmath}

In both the models, we used the {\it MYTZ} component from the {\it MYtorus} set of models \citep{2009MNRAS.397.1549M, 2012MNRAS.423.3360Y} to fit the zeroth-order continuum. This is often called the \say{direct} or \say{transmitted} continuum which is simply a fraction of the intrinsic continuum that leaves the medium without being absorbed or scattered. This energy dependent zeroth-order component ({\it MYTZ}) is used as a multiplicative factor applied to the intrinsic continuum. {\it MYTZ} is purely a line-of-sight quantity and does not depend on the geometry and covering fraction of the out-of-sight material. It includes the equivalent column density ($\rm{N_{H}}$), inclination of the obscuring torus along the line of sight ($\theta_{incl}$) and the redshift of the source. During the spectral fits $\rm{N_{H}}$ and $\theta_{incl}$ were frozen to $10^{25}$ $cm^{-2}$ and $90^{\circ}$ respectively. 

\noindent {\bf Model 2a:} 
During the spectral fitting with this model $\Gamma$ and the normalization for all the transmitted and reflected components were tied together. To model the warm and cold reflectors we used {\it f1*zpo} and {\it xillverCP} respectively. The parameters for these two models were treated in a similar fashion as described in Model 1a. The best fit values obtained from Model 2a are given in Table- \ref{table-8}.

\noindent {\bf Model 2b:} 
Here, $\Gamma$ of the transmitted and scattered components were tied together, but the normalization were varied independently to achieve acceptable fit statistics. For the warm and cold reflectors the model parameters were treated in a similar way as described in Model 1b. Using this model we obtained a better fit statistics than Model 1b with no prominent residues present in the hard energy part (see the right panel of Fig. \ref{figure-7}). The best fit values are given in Table \ref{table-8}. For Model 2b we calculated the flux in three energy bands, i.e, 4$-$10 keV, 10$-$20 keV and 20$-$79 keV. The fluxes obtained for FPMA module ($F^{FPMA}$) are given in Table \ref{table-8}. In the 20 $-$ 79 keV band on epoch D (August, 2014) the source was brighter by about 22\% and 28\% respectively compared to the mean brightness in December 2012 (epoch A, B and C) and February 2015 (epoch E) FPMA spectra. The source brightness again increased in the 20 $-$ 79 keV band on epoch G (August 2017) and it was found to be brighter by about 32\% and 36\% relative to the December 2012 and February 2015 spectra respectively. For all the epochs, the best fit data to model residues are plotted in the right panel of Fig. \ref{figure-8}.

\begin{figure}
	\hspace{-1.5 cm}
	\begin{minipage}{1.00\columnwidth}
		\centering
		\includegraphics[width=\textwidth]{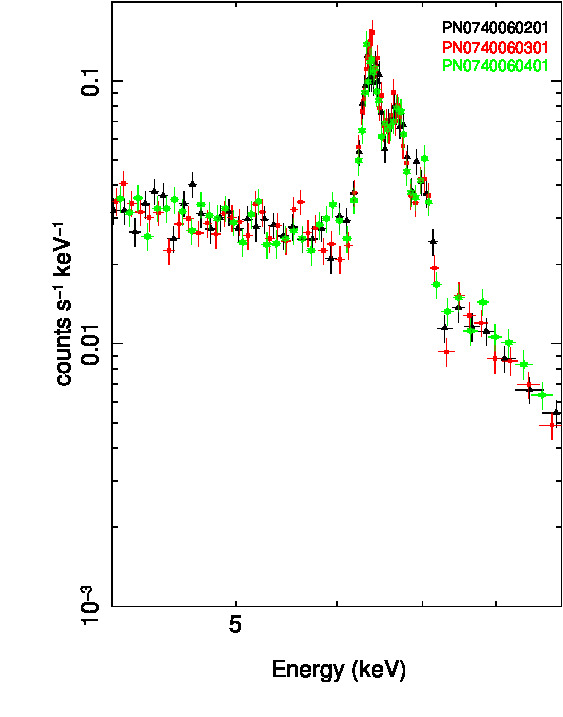}
		\end{minipage}%
\caption{Three 2014 {\it EPIC-PN} spectra plotted together in the 4$-$9 keV band.} \label{figure-10}		
\end{figure}

\begin{figure*}
	\centering
	\begin{minipage}{2.10\columnwidth}
		\centering
		\includegraphics[width=\textwidth]{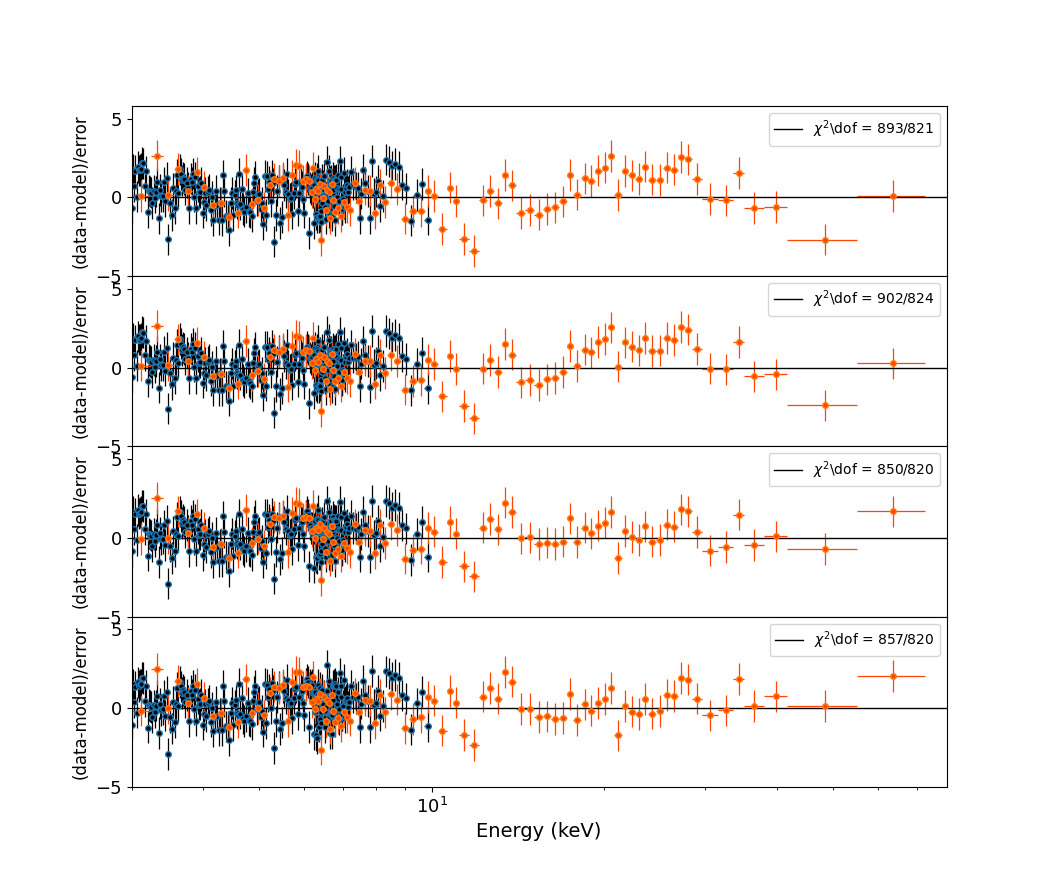}
		\end{minipage}%
\caption{Best fit data to the model ratio for {\it constant*phabs*zphabs*(cutoffpl+pexrav+zgauss(8))} (top panel) ; {\it constant*phabs*zphabs*(f1*cutoffpl+pexrav+zgauss(8))} (2nd panel from the top) ; {\it constant*phabs*zphabs*(MYTZ*cutoffpl+f1*cutoffpl+pexrav+zgauss(8))} (3rd panel from the top) and {\it constant*phabs*zphabs*(zpo*MYTZ+f1*zpo+xillverCP+zgauss(6))} (bottom panel) to the {\it XMM-Newton and NuSTAR} (epoch D FPMA) spectra in the 3$-$79 keV band.} \label{figure-13}
\end{figure*}

\begin{figure*}
	\centering
	\begin{minipage}{1.03\columnwidth}
		\centering
		\includegraphics[width=\textwidth]{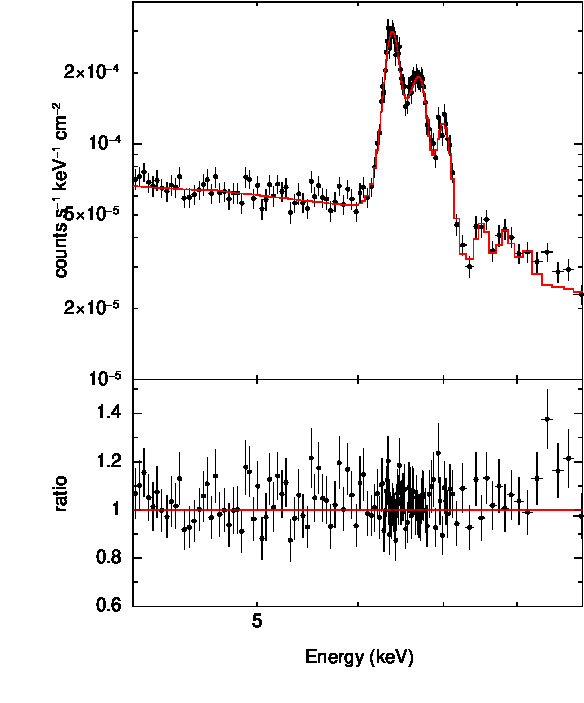}
		\end{minipage}%
	\begin{minipage}{1.09\columnwidth}
		\centering
		\includegraphics[width=\textwidth]{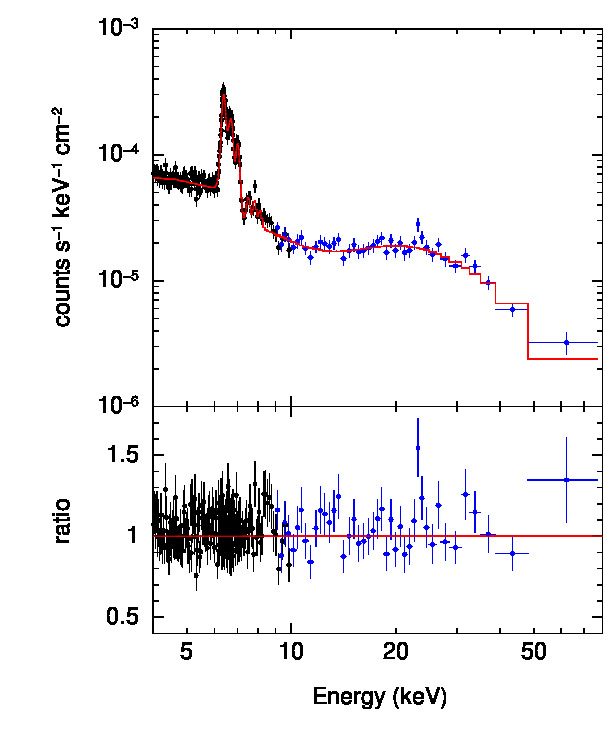}
		\end{minipage}
\caption{Left panel: Best fit {\it EPIC-PN} combined spectra in the 4$-$9 keV band. Right panel: {\it XMM-Newton and NuSTAR} (epoch D FPMA) joint best fit spectra in the 4$-$79 keV band.} \label{figure-11}		
\end{figure*}

\begin{table}
\caption{Best fit line energies along with normalization. Here, the line energy (E) is in keV and the normalization ($N_{E}$) is in units of $10^{-5}$ photons keV$^{-1}$ cm$^{-2}$ s$^{-1}$} \label{table-7}
\begin{tabular}{ccccccr}
\hline
Parameter  & &  line  & & Value \\
\hline
E1    & & Fe Be-like K$\alpha$ & &  6.60$^{+0.03}_{-0.03}$     \\
$N_{E1}$ & &  & &  1.85$^{+0.47}_{-0.44}$  \\
E2  & &  Fe He-like K$\alpha$  & & 6.75$^{+0.02}_{-0.02}$   \\
$N_{E2}$ & &    & & 2.86$^{+0.46}_{-0.50}$  \\
E3       & &  Fe H–like K$\alpha$ & &  7.02$^{+0.01}_{-0.02}$   \\
$N_{E3}$ & &  & & 1.35$^{+0.19}_{-0.18}$ \\
E4       & &  Ni K$\alpha$  & & 7.53$^{+0.03}_{-0.03}$   \\
$N_{E4}$ & &    & & 0.51$^{+0.15}_{-0.15}$  \\
E5       & &  Ni ionized He-like K$\alpha$ & & 7.86$^{+0.04}_{-0.04}$   \\
$N_{E5}$ & &    & & 0.46$^{+0.15}_{-0.15}$  \\
E6       & &  Ni K$\beta$ & & 8.15$^{+0.08}_{-0.08}$   \\
$N_{E6}$ & &    & & 0.28$^{+0.14}_{-0.14}$  \\
\hline
\end{tabular}
\end{table}

\begin{table*}
\caption{Results of Model 1a, Model 1b, Model 2a and Model 2b fits to the simultaneous {\it NuSTAR} FPMA $-$ FPMB spectra. The $\rm{kT_{e}}$ and the line energies (E1, E2, E3 and E4) are in units of keV. Column densities ($N_{H}$) are in unit of $cm^{-2}$. Flux (F) is expressed in unit of $10^{-11}$ erg $cm^{-2}$ $s^{-1}$. Normalization of components (N) in different models at 1 keV are in units of photons $keV^{−1}$ $cm^{−2}$ $s^{−1}$. Parameters with the star (*) mark represent the frozen values.} \label{table-8}
\centering
\begin{tabular}{cccccccccc}
\hline
Model & Parameter & epoch A & epoch B & epoch C & epoch D & epoch E & epoch F & epoch G & epoch H  \\
\hline\hline
1a & $\Gamma$ & 1.33$^{+0.03}_{-0.03}$ & 1.30$^{+0.03}_{-0.03}$ & 1.32$^{+0.05}_{-0.06}$ & $<$1.21 & 1.32$^{+0.03}_{-0.03}$ & $<$1.22 & $<$1.21 & $<$1.22   \\

& $\rm{A_{Fe}}$ & 4.79$^{+0.71}_{-0.37}$ & 5.00* & 4.87$^{+1.50}_{-0.57}$ & 4.38$^{+0.30}_{-0.32}$ &  4.34$^{+0.36}_{-0.36}$ & 4.93$^{+0.47}_{-0.35}$ & 4.46$^{+0.30}_{-0.30}$ & 4.40$^{+0.30}_{-0.30}$ \\

& $\rm{kT_{e}}$ & 7.60$^{+0.47}_{-0.41}$ & 7.48$^{+0.35}_{-0.32}$ & 8.04$^{+0.78}_{-0.69}$ & 7.44$^{+0.28}_{-0.27}$ &  6.82$^{+0.36}_{-0.32}$ & 7.26$^{+0.25}_{-0.29}$ & 7.60$^{+0.29}_{-0.28}$ & 7.29$^{+0.29}_{-0.28}$ \\

& N $\times(10^{-4})$ & 1.31$^{+0.10}_{-0.10}$ & 1.35$^{+0.07}_{-0.08}$ & 1.33$^{+0.17}_{-0.15}$ & 1.60$^{+0.08}_{-0.08}$ & 1.50$^{+0.11}_{-0.10}$ & 1.43$^{+0.08}_{-0.08}$ & 1.61$^{+0.09}_{-0.09}$ & 1.53$^{+0.09}_{-0.09}$ \\

& p1 & 1.28$^{+0.24}_{-0.20}$ & 1.23$^{+0.16}_{-0.14}$ & 1.31$^{+0.35}_{-0.31}$ & 0.70$^{+0.11}_{-0.10}$ &  0.93$^{+0.18}_{-0.16}$ & 0.76$^{+0.10}_{-0.11}$ & 0.66$^{+0.11}_{-0.10}$ & 0.69$^{+0.11}_{-0.10}$ \\

& $\chi^2/dof$ & 468/482 & 437/421 & 210/199 & 483/468 & 436/452 & 473/414 & 528/449 & 479/429 \\

& E1  & 6.75$^{+0.03}_{-0.03}$ & 6.79$^{+0.03}_{-0.03}$ & 6.79$^{+0.05}_{-0.05}$ & 6.77$^{+0.03}_{-0.03}$ & 6.78$^{+0.03}_{-0.03}$ & 6.77$^{+0.03}_{-0.03}$ & 6.80$^{+0.03}_{-0.03}$ & 6.79$^{+0.03}_{-0.03}$    \\
& $N_{E1}$ $\times(10^{-5})$ & 3.82$^{+0.42}_{-0.47}$ & 3.50$^{+0.45}_{-0.42}$ & 3.61$^{+0.37}_{-0.69}$ & 3.39$^{+0.46}_{-0.43}$ & 3.40$^{+0.45}_{-0.42}$ & 2.64$^{+0.44}_{-0.41}$ & 2.88$^{+0.42}_{-0.40}$ & 3.39$^{+0.44}_{-0.43}$ \\
& E2  & 7.59$^{+0.09}_{-0.09}$ & 7.58$^{+0.16}_{-0.13}$ & 7.62$^{+0.16}_{-0.15}$ & 7.45$^{+0.18}_{-0.16}$ & 7.51$^{+0.14}_{-0.18}$ & 7.61$^{+0.10}_{-0.09}$ & 7.66$^{+0.15}_{-0.13}$ & 7.55$^{+0.09}_{-0.10}$  \\
& $N_{E2}$ $\times(10^{-5})$ & 0.81$^{+0.21}_{-0.22}$ & 0.55$^{+0.21}_{-0.21}$ & 0.80$^{+0.41}_{-0.44}$ & 0.46$^{+0.23}_{-0.24}$ & 0.55$^{+0.25}_{-0.28}$ & 0.62$^{+0.21}_{-0.21}$ & 0.56$^{+0.21}_{-0.20}$ & 0.62$^{+0.23}_{-0.23}$ \\
& E3  & 8.19$^{+0.12}_{-0.16}$ & 8.46$^{+0.17}_{-0.17}$ & 8.07$^{+0.16}_{-0.15}$ & 7.96$^{+0.11}_{-0.11}$ & 7.95$^{+0.20}_{-0.17}$ & 8.33$^{+0.13}_{-0.14}$ & 8.03$^{+0.23}_{-0.21}$ & 8.08$^{+0.08}_{-0.09}$  \\
& $N_{E3}$ $\times(10^{-5})$ & 0.45$^{+0.19}_{-0.19}$ & 0.48$^{+0.19}_{-0.19}$ & 0.74$^{+0.41}_{-0.41}$ & 0.65$^{+0.22}_{-0.23}$ & 0.43$^{+0.26}_{-0.25}$ & 0.38$^{+0.19}_{-0.18}$ & 0.55$^{+0.23}_{-0.27}$ & 0.56$^{+0.21}_{-0.21}$ \\

& E4  & 8.77$^{+0.11}_{-0.12}$ & 8.75$^{+0.10}_{-0.11}$ & - & 8.63$^{+0.18}_{-0.23}$ & 8.87$^{+0.17}_{-0.36}$ & 9.12$^{+0.19}_{-0.18}$ & 9.00$^{+0.14}_{-0.14}$ & 9.00$^{+0.26}_{-0.29}$  \\
& $N_{E4}$ $\times(10^{-6})$ & 3.69$^{+1.74}_{-1.79}$ & 3.28$^{+1.20}_{-1.25}$ & - & 2.79$^{+1.79}_{-1.79}$ & 3.84$^{+1.72}_{-1.72}$ & 2.15$^{+1.69}_{-1.63}$ & 4.19$^{+1.80}_{-1.79}$ & 2.75$^{+1.76}_{-1.76}$ \\

& $\rm{C_{FPMA/FPMB}}$ & 1.04$^{+0.03}_{-0.03}$ & 1.03$^{+0.03}_{-0.03}$ & 1.01$^{+0.05}_{-0.04}$ & 1.01$^{+0.03}_{-0.03}$ & 1.02$^{+0.03}_{-0.03}$ & 1.00$^{+0.03}_{-0.03}$ & 1.01$^{+0.03}_{-0.03}$ & 0.98$^{+0.03}_{-0.03}$ \\
\hline
1b & $\Gamma$ & 1.34$^{+0.05}_{-0.03}$ & 1.26$^{+0.08}_{-0.06}$ & $<$1.27 & $<$1.24 & 1.31$^{+0.04}_{-0.05}$ & $<$1.26 & $<$1.22 & $<$1.26   \\

& $\rm{A_{Fe}}$ & $<$7.05 & $<$6.31 & $>$7.29 & 5.08$^{+1.31}_{-0.23}$ & 5.83$^{+1.86}_{-1.00}$ & 6.82$^{+0.95}_{-1.78}$ & 5.50$^{+0.69}_{-0.59}$ & 5.53$^{+1.52}_{-0.68}$ \\

& $\rm{kT_{e}}$ & 9.10$^{+0.13}_{-0.16}$ & 8.95$^{+0.16}_{-0.20}$ & 9.38$^{+0.18}_{-0.18}$ & 8.75$^{+0.16}_{-0.12}$ &  8.69$^{+0.22}_{-0.25}$ & 8.68$^{+0.17}_{-0.23}$ & 8.77$^{+0.13}_{-0.14}$ & 8.78$^{+0.16}_{-0.16}$ \\

& $N_{\it xillverCP_{warm}}$ $\times(10^{-5})$ & 2.43$^{+0.24}_{-0.34}$ & 2.29$^{+0.36}_{-0.36}$ & 2.68$^{+0.15}_{-0.49}$ & 2.03$^{+0.30}_{-0.16}$ & 1.90$^{+0.35}_{-0.26}$ & 1.95$^{+0.18}_{-0.35}$ & 1.99$^{+0.15}_{-0.19}$ & 2.02$^{+0.31}_{-0.23}$ \\

& $N_{\it xillverCP_{cold}}$ $\times(10^{-4})$ & 1.03$^{+0.08}_{-0.06}$ & 1.12$^{+0.09}_{-0.10}$ & 1.00$^{+0.11}_{-0.08}$ & 1.38$^{+0.07}_{-0.09}$ & 1.18$^{+0.10}_{-0.10}$ & 1.22$^{+0.10}_{-0.08}$ & 1.39$^{+0.07}_{-0.07}$ & 1.27$^{+0.10}_{-0.09}$ \\

& $\chi^2/dof$ & 504/483 & 474/420 & 213/199 & 545/468 & 480/452 & 545/414 & 600/449 & 545/431 \\
\hline
2a & $\Gamma$ & 1.32$^{+0.03}_{-0.03}$ & 1.30$^{+0.04}_{-0.04}$ & 1.32$^{+0.05}_{-0.05}$ & $<$1.22 & 1.32$^{+0.03}_{-0.03}$ & $<$1.22 & $<$1.21 & $<$1.22   \\

& $\rm{A_{Fe}}$ & 4.78$^{+0.70}_{-0.37}$ & 5.50$^{+1.14}_{-0.77}$ & 4.77$^{+1.47}_{-0.60}$ & 4.43$^{+0.31}_{-0.32}$ &  4.32$^{+0.38}_{-0.37}$ & 5.00$^{+0.63}_{-0.37}$ & 4.56$^{+0.31}_{-0.32}$ & 4.33$^{+0.37}_{-0.34}$ \\

& $\rm{kT_{e}}$ & 7.39$^{+0.41}_{-0.69}$ & 7.80$^{+0.49}_{-0.58}$ & 7.28$^{+0.86}_{-0.62}$ & 7.45$^{+0.30}_{-0.29}$ &  6.35$^{+0.46}_{-0.34}$ & 7.51$^{+0.33}_{-0.37}$ & 7.43$^{+0.35}_{-0.55}$ & 6.80$^{+0.51}_{-0.35}$ \\

& p1 & 0.79$^{+0.69}_{-0.21}$ & 1.34$^{+0.20}_{-0.18}$ & 0.83$^{+0.53}_{-0.35}$ & 0.74$^{+0.11}_{-0.10}$ &  0.53$^{+0.25}_{-0.19}$ & 0.88$^{+0.11}_{-0.11}$ & 0.64$^{+0.12}_{-0.31}$ & 0.72$^{+0.10}_{-0.11}$ \\

& $\chi^2/dof$ & 467/482 & 421/420 & 205/199 & 480/468 & 431/452 & 449/414 & 521/449 & 476/431 \\
\hline
2b & $\Gamma$ & 1.35$^{+0.09}_{-0.07}$ & 1.50$^{+0.08}_{-0.08}$ & 1.49$^{+0.13}_{-0.13}$ & 1.31$^{+0.06}_{-0.04}$ & 1.37$^{+0.03}_{-0.03}$ & 1.35$^{+0.04}_{-0.03}$ & 1.32$^{+0.06}_{-0.05}$ & 1.34$^{+0.05}_{-0.05}$   \\

& $\rm{A_{Fe}}$ & 6.09$^{+2.43}_{-1.44}$ & 4.39$^{+0.76}_{-0.70}$ & 4.48$^{+1.63}_{-0.95}$ & 4.98$^{+0.56}_{-0.63}$ &  5.00* & 5.00$^{+0.71}_{-0.41}$ & 4.43$^{+0.72}_{-0.68}$ & 4.31$^{+0.67}_{-0.66}$ \\

& $\rm{kT_{e}}$ & 8.97$^{+0.22}_{-0.30}$ & 8.51$^{+0.49}_{-0.82}$ & 9.13$^{+0.63}_{-0.98}$ & 8.76$^{+0.15}_{-0.39}$ &  8.55$^{+0.17}_{-0.16}$ & 8.57$^{+0.17}_{-0.32}$ & 8.46$^{+0.39}_{-0.66}$ & 8.30$^{+0.45}_{-0.72}$ \\

& $\chi^2/dof$ & 481/481 & 433/419 & 211/198 & 507/467 & 464/452 & 495/415 & 529/448 & 496/430 \\

& $F_{4-10}^{FPMA}$ & 0.42 & 0.42 & 0.43 & 0.41 & 0.42 & 0.38 & 0.39 & 0.40 \\
& $F_{10-20}^{FPMA}$ & 0.45 & 0.45 & 0.48 & 0.49 & 0.45 & 0.42 & 0.48 & 0.47 \\
& $F_{20-79}^{FPMA}$ & 2.23 & 2.50 & 2.55 & 3.09 & 2.24 & 2.92 & 3.54 & 2.97 \\
\hline
\end{tabular}
\end{table*}

\subsection{XMM-Newton \& NuSTAR Joint fit }
Fitting the {\it NuSTAR} spectra alone could not handle the line emission profiles present in the source spectrum properly. To model the lines we used three {\it XMM-Newton EPIC PN} spectra taken in 2014 along with the {\it NuSTAR} FPMA data. Use of {\it XMM-Newton} data jointly with {\it NuSTAR}, the observations of which are not simultaneous requires the source to be non-variable. We show in Fig. \ref{figure-10} all three {\it XMM-Newton PN} spectra taken in 2014. This figure indicates that the source has not shown any noticeable variation in line and continuum flux. Also, in all the eight epochs of {\it NuSTAR} data accumulated over a period of 5 years, no variation in the soft band (4$-$10 keV) is observed (see Table \ref{table-4} and Fig. \ref{figure-2}). Therefore, it is not inappropriate to jointly model the {\it NuSTAR} and {\it XMM-Newton} observations.  We combined the three {\it XMM-Newton EPIC PN} spectra together using the task {\it epicspeccombine} and then binned the spectra with 25 counts/bin using the task {\it specgroup}. 

We carried out joint model fits to the 3-10 keV {\it XMM-Newton} 2014 combined spectra with the 3-79 keV epoch D {\it NuSTAR} spectrum. To account for both warm and cold reflection, several models were tried. Firstly, we used a {\it cutoffpl} to model the warm reflector that did not take into account Compton down scattering. For modelling the cold reflector we used {\it pexrav} \citep{10.1093/mnras/273.3.837} with R=-1. We obtained the best fit values of 1.59$^{+0.09}_{-0.09}$, 95$^{+100}_{-47}$ and 11.22$^{+3.51}_{-2.70}$ for $\Gamma$, $\rm{E_{cut}}$ and $\rm{A_{Fe}}$ respectively with a $\chi^2$ of 893 for 821 degrees of freedom. Using this model we got an acceptable fit statistics with a mild hump between 20 $-$ 30 keV energy range (see top panel of Fig. \ref{figure-13}). We replaced the {\it cutoffpl} with {\it xillver} with a fixed $\log\xi$ = 4.7 to account for the Compton down scattering. We obtained $\Gamma$ = 1.54$^{+0.04}_{-0.04}$, $\rm{E_{cut}}$ = 94$^{+26}_{-19}$ and $\rm{A_{Fe}}$ $>$ 9.15. This fit produced a $\chi^2$ of 884 for 821 degrees of freedom. We then replaced {\it xillver} with {f1*cutoffpl} (for f1, see Equation 3) to model the warm reflector and obtained $\Gamma$ = 1.62$^{+0.11}_{-0.11}$, $\rm{E_{cut}}$ = 97$^{+149}_{-39}$ and $\rm{A_{Fe}}$ = 11.50$^{+4.54}_{-3.11}$ with a $\chi^2$ of 902 for 824 degrees of freedom. With or without the inclusion of Compton down scattering in the warm reflector we obtained similar set of best fit values and a hump in the data to model residue plot near 25 keV (see first two panels of Fig. \ref{figure-13}). We thus conclude that inclusion of Compton down scattering in the warm reflection has insignificant effect on the derived parameters.

The spectrum of the Compton-Thick AGN is refection dominated, however, there is a finite probability for the primary emission to be transmitted (above $>$10 keV) through the Compton thick absorber. Thus, we modified our model to take into account the transmitted primary emission by including the ({\it MYTZ*cutoffpl}) component into the previously described two reflector models in which the cold reflection was modelled using {\it pexrav} with R=-1. For the warm reflector we first used {\it xillver} with R=-1 and $log\xi$ = 3.1. For the Compton absorber along the line of sight we assumed a column density of $10^{25}$ $cm^{-2}$ with an inclination of $90^{\circ}$ \citep{Bauer_2015}. We obtained $\Gamma$ $<$1.28, $\rm{E_{cut}}$ = 16$^{+2}_{-1}$ and $\rm{A_{Fe}}$ = $>$7.41 with a $\chi^2$ of 847 for 820 degrees of freedom. The fit statistics now improved by $\Delta\chi^2$ = 37 over the reduction of 1 degree of freedom and the hump around 20 $-$ 30 keV band was also taken care of. We obtained a similar fit with $\Gamma$ = 1.37$^{+0.18}_{-0.16}$, $\rm{E_{cut}}$ = 18$^{+4}_{-3}$ and $\rm{A_{Fe}}$ = 5.68$^{+4.20}_{-2.41}$ with a $\chi^2$ of 850 for 820 degrees of freedom from using {\it f1*cutoffpl} (see Equation 3) as a warm reflector. Inclusion of the transmitted primary emission with the two reflector model described the source spectra well with a harder photon index and lower $\rm{E_{cut}}$ with no prominent features present in the data to model ratio plot at high energies (see 3rd panel of Fig. \ref{figure-13}). Previously also, NGC 1068 X-ray spectrum was modelled with a flatter photon index. From a joint analysis of {\it XMM-Newton} epoch E spectrum with the epoch D {\it NuSTAR} spectrum \cite{2021MNRAS.506.4960H} reported $\Gamma$ =  1.21$^{+0.13}_{-0.07}$. From modelling the 0.5$-$10 keV {\it ASCA} observation \cite{1994PASJ...46L..71U} found a best fit $\Gamma$ of 1.28$\pm$0.14. Replacing {\it pexrav} with {\it xillverCP} to model the cold reflector with R=-1 and $log\xi$ = 0.0 we obtained $\Gamma$ = 1.26$^{+0.03}_{-0.03}$, $\rm{kT_{e}}$ = 8.59$^{+0.40}_{-0.37}$ and $\rm{A_{Fe}}$ = 4.02$^{+0.36}_{-0.33}$ with a $\chi^2$ = 857 for 820 degrees of freedom. The data to the best fit model residues are given in Fig. \ref{figure-13}.

To estimate $\rm{kT_{e}}$ we used Model 2b, as described earlier for the {\it NuSTAR} spectral fit alone. Here the {\it NuSTAR}  FPMA spectra for all the epochs in the 9 $-$ 79 keV along with the combined 2014 {\it XMM-Newton} data in the 4 $-$ 10 keV band were used to take care of the line emission region carefully. We used six Gaussian components to model all the ionized and neutral lines present in the spectra. The line energies and the normalization were kept free during the fitting while the line widths were frozen to 0.01 keV. The 
self-consistent {\it $xillverCP_{cold}$} model took care of the neutral Fe k$\alpha$ ($\sim$ 6.4 keV) and k$\beta$ ($\sim$ 7.08 keV) lines, wherein the first three Gaussian components were used to model the ionized Fe lines at energies $\sim$ 6.57 keV (Fe Be-like K$\alpha$), 6.67 keV (Fe He-like K$\alpha$) and 6.96 keV (Fe H–like K$\alpha$). The neutral Ni k$\alpha$ ($\sim$ 7.47 keV), Ni k$\beta$ ($\sim$ 8.23 keV) and Ni ionized He-like K$\alpha$ ($\sim$ 7.83 keV) were taken care by the other three Gaussian components. As seen from the left panel of Fig. \ref{figure-11} we did not find any prominent residue in the 4$-$9 keV band. All the best fitted line energies and the normalization are given in Table \ref{table-7}. The best fit model parameters along with their corresponding errors are given in Table \ref{table-9}. We found that this joint fit produced similar best fit values as obtained from {\it NuSTAR} fit alone. The best fit model to the data (the combined {\it EPIC PN} data and {\it NuSTAR} FPMA epoch D spectra) along with the residue is shown in the right panel of Fig \ref{figure-11}.

\begin{table*}
\caption{Results of the analysis of the Model 2b fit to the {\it XMM-Newton} and {\it NuSTAR} FPMA spectra in the 4 $-$ 79 keV energy band. $\rm{kT_{e}}$ is in unit of keV and column densities ($N_{H}$) are in units of $cm^{-2}$. Normalization of components (N) at 1 keV is in unit of photons $keV^{−1}cm^{−2}s^{−1}$.} \label{table-9}
\centering
\begin{tabular}{p{0.12\linewidth}p{0.08\linewidth}p{0.08\linewidth}p{0.08\linewidth}p{0.08\linewidth}p{0.08\linewidth}p{0.08\linewidth}p{0.08\linewidth}p{0.08\linewidth}}
\hline
Parameter & epoch A & epoch B & epoch C & epoch D & epoch E & epoch F & epoch G & epoch H  \\
\hline\hline
$N_{H}^{ztbabs}$ & 9.78$^{+2.92}_{-2.96}$ & 9.19$^{+2.73}_{-2.75}$ & 7.09$^{+5.04}_{-4.19}$ & 9.57$^{+2.39}_{-2.60}$ & 9.16$^{+2.58}_{-2.72}$ & 9.72$^{+2.37}_{-2.55}$ & 9.51$^{+2.75}_{-3.15}$ & 8.90$^{+2.64}_{-2.88}$  \\

$\Gamma$ & 1.29$^{+0.07}_{-0.08}$ & 1.33$^{+0.15}_{-0.07}$ & 1.48$^{+0.11}_{-0.25}$ & $<$1.32 & 1.33$^{+0.10}_{-0.06}$ & 1.28$^{+0.07}_{-0.05}$ & 1.26$^{+0.07}_{-0.10}$ & 1.30$^{+0.07}_{-0.07}$  \\

$\rm{A_{Fe}}$ & 6.08$^{+3.87}_{-1.59}$ & 4.89$^{+2.89}_{-1.51}$ & 3.09$^{+4.96}_{-1.08}$ & 4.19$^{+1.44}_{-0.93}$ & 4.75$^{+2.19}_{-0.63}$ & 4.99$^{+1.70}_{-1.10}$ & 3.90$^{+1.33}_{-1.10}$ & 3.81$^{+1.27}_{-1.01}$  \\

$\rm{kT_{e}}$ & 8.69$^{+0.28}_{-0.33}$ & 8.62$^{+0.33}_{-0.49}$ & 8.07$^{+0.59}_{-0.75}$ & 8.59$^{+0.30}_{-0.36}$ & 8.60$^{+0.34}_{-0.58}$ & 8.68$^{+0.26}_{-0.38}$ & 8.55$^{+0.25}_{-0.41}$ & 8.48$^{+0.35}_{-0.51}$ \\

$\chi^2/dof$ & 592/584 & 576/565 & 516/492 & 620/586 & 590/574 & 613/569 & 630/584 & 585/578 \\

$\rm{C_{XMM/NuSTAR}}$ & 0.96$^{+0.08}_{-0.08}$ & 0.95$^{+0.08}_{-0.07}$ & 0.94$^{+0.11}_{-0.11}$ & 0.94$^{+0.08}_{-0.08}$ & 0.93$^{+0.08}_{-0.08}$ & 0.98$^{+0.08}_{-0.08}$ & 0.93$^{+0.08}_{-0.08}$ & 0.97$^{+0.08}_{-0.09}$ \\
\hline
\end{tabular}
\end{table*}

\section{Summary}
In this work, we carried out spectral and timing  analysis of eight epochs of {\it NuSTAR}  
observations performed between December 2012 and November 2017 probing time scales within
epochs and between epochs that spans about 5 years. The timing analysis of the six {\it XMM-Newton} observations between July 2000 and February 2015 was also performed. We also carried out the spectral analysis of the 2014 combined {\it XMM-Newton EPIC PN} and {\it NuSTAR} FPMA data jointly. The results of this work are summarized below
\begin{enumerate}
\item We found the source not to show flux variation within each of the eight epochs of {\it NuSTAR} observation.
\item Between epochs, that span the time-scales from 2012 to 2017, we found variation
in the source. Here too, the source did not show variation in the soft energy range.
As in agreement with the earlier results by \cite{2016MNRAS.456L..94M} and \cite{2020MNRAS.492.3872Z}, we also found that the observed variations is only due to variation in the energy range beyond 20 keV. This too was noticed in Epoch D (August 2014) and Epoch G (August 2017), when the brightness of the source beyond 20 keV was higher by about 20\% and 30\%  respectively relative to the three {\it NuSTAR} observations in the year 2012.
\item From timing analysis, we observed no correlation of spectral variation (hardness ratio) with brightness.
\item Fitting physical models to the observed data we could determine the  temperature
of the corona in NGC 1068 with values ranging from  8.46$^{+0.39}_{-0.66}$ keV and 9.13$^{+0.63}_{-0.98}$ keV. However, we found no variation in the temperature of the corona during the 8 epochs of observations that span a duration of about 5 years.

\item From the timing analysis of six {\it XMM-Newton EPIC PN} data we found no significant flux variation both in between and within epochs of observation in the hard band. In the soft band too we found the source did not show any significant flux variation within epochs but it was brighter in epoch B compared to epoch A.
\item The combined spectral fit of {\it XMM-Newton} and {\it NuSTAR} data provided results that are in agreement with those obtained by model fits to the {\it NuSTAR} data alone. 
\end{enumerate}

In NGC 1068, we did not find evidence for variation in the temperature of the corona from analysis of data that span more than five years. This is evident from the best fit values of $\rm{kT_{e}}$ from Table \ref{table-8}. Also, the results from various models are found to be similar. The values of $\rm{kT_{e}}$ found for NGC 1068 also lie in the range of $\rm{kT_{e}}$ found in other AGN. Measurements of $\rm{E_{cut}}$ are available for
a large number of AGN that includes both Seyfert 1 and Seyfert 2 type. However, studies on the variation of $\rm{E_{cut}}$ or $\rm{kT_{e}}$ are limited to less
than a dozen AGN \citep{2014ApJ...794...62B, 2015A&A...577A..38U, 2016MNRAS.463..382U, 2016MNRAS.456.2722K, 2017ApJ...836....2Z, 2018ApJ...863...71Z, 2020MNRAS.492.3041B, 2021MNRAS.502...80K, Barua_2021, 2022A&A...662A..78P}. Even in sources where $\rm{E_{cut}}$/$\rm{kT_{e}}$ variations are known, the correlation of the variation of $\rm{kT_{e}}$ with various physical
properties of the sources are found to be varied among sources \citep{2020MNRAS.492.3041B, Barua_2021, 2021MNRAS.502...80K, 2022A&A...662A..78P}. These limited
observations do indicate that we do not yet understand the complex corona of AGN including its geometry and composition. Investigation of this
kind needs to be extended for many AGN to better constrain the nature of corona.

\section*{Acknowledgements}
We thank the anonymous referee for his/her comments that helped us in correcting an error in the analysis. We also thank Drs. Ranjeev Misra and Gulab Dewangan for discussion on spectral fits to the data. We thank the {\it NuSTAR} Operations, Software and Calibration teams for support with the execution and analysis of these observations. This research has made use of the {\it NuSTAR} Data Analysis Software (NuSTARDAS) jointly developed by the ASI Science Data Center (ASDC, Italy) and the California Institute of Technology (USA). This research has made use of data and/or software provided by the High Energy Astrophysics Science Archive Research Center (HEASARC), which is a service of the Astrophysics Science Division at NASA/GSFC. VKA  thanks GH, SAG; DD, PDMSA and Director, URSC for encouragement and continuous support to carry out this research.

\section*{Data Availability}

All data used in this work are publicly available in the {\it NuSTAR} (\url{https://heasarc.gsfc.nasa.gov/docs/nustar/nustar_archive.html}) and {\it XMM-Newton} (\url{http://nxsa.esac.esa.int}) science archive.



\bibliographystyle{mnras}
\bibliography{example} 



\bsp	
\label{lastpage}
\end{document}